\documentclass[12pt]{article}
\usepackage{amssymb,amsmath,latexsym,graphicx, mathrsfs, hyperref, bm}

\begin{document}
\title{On a reduction of the generalized \\ 
Darboux-Halphen system \author{ Sumanto Chanda$^1$, Sarbarish Chakravarty$^2$, Partha Guha$^{1,3}$}}

\maketitle
\thispagestyle{empty}

\begin{minipage}{0.35\textwidth}
\begin{flushleft}
\textit{\small $ ^1$S.N. Bose National Centre \\ for Basic Sciences} \\
\textit{\small JD Block, Sector-3, Salt Lake, Calcutta-700098, INDIA.} \\ \texttt{\small sumanto12@boson.bose.res.in, partha@bose.res.in}
\end{flushleft}
\end{minipage}
\begin{minipage}{0.3 \textwidth}
\begin{center} \large
\textit{\small $ ^2$ Department of Mathematics} \\
\textit{\small University of Colorado, Colorado Springs, CO 80918, USA. } \\
\texttt{\small schakrav@uccs.edu}
\end{center}
\end{minipage}
\begin{minipage}{0.3\textwidth}
\begin{flushright}
\textit{\small $ ^3$IHES, Le Bois-Marie 35\\ 
route de Chartres 91440, \\ Bures-sur-Yvette, \\ France.} \\ \texttt{\small guha@ihes.fr}
\end{flushright}
\end{minipage}
\bigskip

{\bf{Keywords :}} Darboux-Halphen, Self-Dual Yang-Mills, Bianchi-IX, Lax representation, 

Hypergeometric function

\date{}


\abstract{The equations for the general Darboux-Halphen system obtained
as a reduction of the self-dual Yang-Mills can be transformed
to a third-order system which resembles the classical Darboux-Halphen
system with a common additive terms. It is shown that the
transformed system can be further reduced to a constrained
non-autonomous, non-homogeneous dynamical system. This dynamical
system becomes homogeneous for the classical
Darboux-Halphen case, and was studied in the context of
self-dual Einstein's equations for Bianchi IX metrics. 
A Lax pair and Hamiltonian for this reduced system is derived 
and the solutions for 
the system are prescribed in terms of hypergeometric functions.}

\setcounter{page}{1}
\numberwithin{equation}{section}

\section{Introduction} \label{sec:sec1}

The Darboux-Halphen differential equations often referred to as 
the classical Darboux-Halphen (DH) system
\begin{equation}
\label{DH} 
\dot{\omega}_i = \omega_j\omega_k-\omega_i\left(\omega_j+\omega_k\right) \,,  
\qquad i \neq j \neq k = 1, 2, 3, \, cyclic\,, \qquad \dot{} := \frac{d}{dt}\,,
\end{equation}
was originally formulated by Darboux \cite{Da} and subsequently 
solved by Halphen \cite{Ha}. 
The general solution to equation 
(\ref{DH}) may be expressed in terms of the elliptic modular function. 
In fact Halphen related the DH equation in terms of the 
null theta functions.

The system (\ref{DH}) has found applications in mathematical 
physics in relation to magnetic monopole dynamics 
\cite{AH}, self dual Einstein equations \cite{GP, Hi}, topological field 
theory \cite{Du} and reduction of self-dual Yang-Mills (SDYM) 
equations~\cite{ACC}. Recently in \cite{cgr}, the DH system was
reviewed from the perspective of the self-dual Bianchi-IX
metric and the SDYM field equations, describing
a gravitational instanton in the former case, and a
Yang-Mills instanton in the latter. All systems related
to the DH system such as Ramanujan and
Ramamani system were covered, as well as aspects of
integrability of the DH system. 

Ablowitz et al \cite{ACH, ACH1} 
studied the reduction of the SDYM equation with an infinite-dimensional Lie 
algebra to a $3 \times 3$ matrix differential equation. This work
led to a generalized Darboux-Halphen (gDH) system which differs from
the DH system by a common additive term.
The gDH system was also solved 
originally by Halphen~\cite{Ha1} in terms of
general hypergeometric functions and whose general solution admits 
movable natural barriers which can be densely branched. 

In this article, we discuss certain aspects related to the
integrability of the gDH system. Some of these features were
implicit in the original formulation of the system but were
never made concrete. Specifically, we show that it is possible
to derive naturally from the gDH system yet another reduced 
system of equations which satisfy a constraint. This constrained
system resembles a non-autonomous Euler equation similar to that
derived by Dubrovin~\cite{Du1} but with non-homogeneous terms.
Furthermore, we derive a simple Lax pair for the constrained system.
The paper is organized as follows. In Section 2, the gDH system is 
introduced and a constrained system is derived from it. Then the
solutions of both the gDH and the constrained systems are discussed.
In Section 3, we derive following~\cite{ACH1}, the gDH system from 
a ninth-order dynamical system that is obtained as a
reduction of the SDYM field equations equation.
We provide some details in our derivation that were not included
in earlier papers.  
Then we discuss the constrained system in the framework of a fifth-order
system that arise as a special case of the SDYM reduction.
In Section 4, we formulate a Lax pair and a Hamiltonian
for the reduced system introduced in Section 2.

\section{The gDH system}
In this section, we introduce the gDH system for the complex
functions $\omega_i(t)$
\begin{equation}
\label{gendh} 
\dot{\omega}_i = \omega_j\omega_k-\omega_i\left(\omega_j+\omega_k \right)+\tau^2\,, 
\quad i \neq j \neq k=1,2,3, \, cyclic \,.
\end{equation}
The common additive term $\tau^2$ is elaborated as
\begin{equation} \label{diff}
\begin{split}
\tau^2 = &\alpha_1^2 x_2 x_3 + \alpha_2^2 x_3 x_1 + \alpha_3^2 x_1 x_2 \qquad
\text{with} \qquad x_i = \omega_j-\omega_k \,, \\
&i \neq j \neq k, \, cyclic\,, \qquad  x_1+x_2+x_3=0\,, \end{split}
\end{equation}
where $\alpha_i\,, \, i=1,2,3$ are complex constants.
As mentioned in Section 1, the gDH system arises from a particular
reduction of the SDYM equations \cite{ACH, ACH1}. They also appear 
in the study of $SU(2)$-invariant,
hypercomplex four-manifolds~\cite{Hi1}. In Section 3, we will
provide a derivation of the gDH system from the SDYM reductions
following \cite{ACH1}.

In the following, we derive from (\ref{gendh}) a reduced system of differential 
equations which satisfy a constraint.

\subsection{Constrained gDH system}
Note that the variables $x_i$ defined in (\ref{diff}) satisfy
the equations
\begin{equation} \label{xprop1} 
\dot{x}_i = - 2 \omega_i x_i\,, \quad i=1,2,3\,, 
\end{equation}
which are obtained from \eqref{gendh} by taking the difference of the
equations for $\omega_j$ and $\omega_k$. Using (\ref{xprop1}), 
the gDH equations \eqref{gendh} can be re-expressed as follows:
\[\dot{\omega}_i -\frac{\omega_i}{2} \left(\frac{\dot{x}_j}{x_j}+\frac{\dot{x}_k}{x_k} \right)
= \omega_j\omega_k+\tau^2\,. \]
Then by defining new variables $W_i\,, \, i=1,2,3$ via
\begin{equation}
\label{W}
W_i := \frac{\omega_i}{\sqrt{x_jx_k}}\,, \quad i \neq j \neq k, \, cyclic \,,
\end{equation}
one obtains the system
\begin{equation}
\dot{W}_i=x_iW_jW_k+\frac{\tau^2}{\sqrt{x_jx_k}}\,.
\label{Weq}
\end{equation}
It follows from \eqref{Weq} that
\[\sum_{i=1}^3W_i\dot{W}_i=W_1W_2W_3\sum_{i=1}^3x_i -
\frac{\tau^2}{2x_1x_2x_3} \sum_{i=1}^3\dot{x}_i =0 \]
after using \eqref{W}, \eqref{xprop1} and the fact that $x_1+x_2+x_3=0$.
Thus, one finds that the quantity
\[Q := \sum_{i=1}^3W_i^2 = 
\frac{\omega_1^2}{x_2x_3}+\frac{\omega_2^2}{x_1x_3}+\frac{\omega_3^2}{x_1x_2}\]
is a constant. However, the quantity $Q$ is {\it not} a conserved quantity
of \eqref{Weq}, rather $Q=-1$ is an identity which follows from
the definition of the variables $W_i$ in \eqref{W}. Indeed, a 
direct calculation using $x_1+x_2+x_3=0$, shows that
\[ \begin{split}
Q &= \frac{\omega_1^2x_1+\omega_2^2x_2+\omega_3^2x_3}{x_1x_2x_3}
= \frac{\omega_1^2x_1+\omega_2^2x_2-\omega_3^2(x_1+x_2)}{x_1x_2x_3} \\
&= \frac{x_1(\omega_1-\omega_3)(\omega_1+\omega_3) + 
x_2(\omega_2-\omega_3)(\omega_2+\omega_3)}{x_1x_2x_3}    
= \frac{x_1x_2(\omega_2-\omega_1)}{x_1x_2x_3}=- \frac{x_1x_2x_3}{x_1x_2x_3}=-1
\end{split} \]
Therefore, the system in \eqref{Weq} is a reduction of the original gDH system;
the reduced system can be regarded as a third order system for the $W_i$ satisfying
the constraint $Q=-1$. Note that the DH equations \eqref{DH} being a
special case ($\alpha_i=0$) of \eqref{gendh}, also admits the same reduced system
\eqref{Weq} as above but with $\tau=0$.    

{\bf Remark}:\, A third order system similar to \eqref{Weq} 
but without the non-homogeneous term, was 
introduced in~\cite{SChak, Tod} where
the authors derived a family of self-dual, SU(2)-invariant, 
Bianchi-IX metrics obtained from solutions
of a special Painleve-VI equation. In that case, the vanishing 
of the anti-self-dual Weyl tensor and scalar curvature led to a
sixth order system described by the classical 
DH system \eqref{DH} coupled to another third order system. 
The $W_i$ variables represented different quantities in~\cite{SChak, Tod}
although they were defined in the same way as in \eqref{W}.
The quantity $Q$ was a first integral (instead of a number) in that
case, depending on the
initial conditions for the sixth order system. This sixth
order system considered in~\cite{SChak, Tod} also admits a special
reduction to the third order DH system when the metric is
self-dual Einstein. It is this latter case which corresponds to
the homogeneous version of \eqref{Weq} above with $Q=-1$.
 
Next, we discuss the solution of the reduced system via
the solutions of the original gDH system \eqref{gendh}.

\subsection{Solutions}
As mentioned in \hyperref[sec:sec1]{Section 1}, Halphen~\cite{Ha1} solved
the gDH system and expressed its solution in terms of the general
hypergeometric equation. Below we discuss a method of solution
first given by Brioschi~\cite{Brioschi}.

Let us first introduce a function $s(t)$ via the following ratio:
\begin{equation}
\label{ratio} s=\frac{\omega_3-\omega_2}{\omega_1-\omega_2} = - \frac{x_1}{x_3}.
\end{equation}
Taking the derivative of $\ln s$ in \eqref{ratio}
and then using \eqref{xprop1}, the $x_i$ can be written as
\begin{equation}
\label{xs}
x_1 = -\frac12 \frac{\dot s}{s-1}\,, \qquad
x_2=\frac12 \frac{\dot s}{s}\,, \qquad  
x_3=\frac12 \frac{\dot s}{s(s-1)}\,. 
\end{equation}
Using \eqref{xprop1} once more, the gDH variables $\omega_i$ can be
expressed in terms of $s, \dot{s}$ and $\ddot{s}$ as
\begin{equation}
\label{omegas}
\omega_1=-\frac12\frac{d}{dt}\left[\ln\left(\frac{\dot s}{s-1}\right)\right]\,, \qquad
\omega_2=-\frac12\frac{d}{dt}\left[\ln\left(\frac{\dot s}{s}\right)\right]\,, \qquad
\omega_3=-\frac12\frac{d}{dt}\left[\ln\left(\frac{\dot s}{s(s-1)}\right)\right]\,. 
\end{equation}
Substituting the above expressions for $\omega_i$ into the gDH system \eqref{gendh}
yields the following third order equation for $s(t)$
\begin{equation}
\label{schwarzian} 
\frac{\dddot s}{\dot s}-\frac32\left(\frac{\ddot s}{\dot s}\right)^2 
+ \frac{\dot s^2}2 \left[ \frac{1-\alpha_1^2}{s^2}+\frac{1-\alpha_2^2}{(s-1)^2}+ 
\frac{\alpha_1^2+\alpha_2^2 - \alpha_3^2-1}{s(s-1)} \right] \,,
\end{equation}
also known as the Schwarzian equation. Equation \eqref{schwarzian}
can be {\it linearized} in terms of the hypergeometric equation as follows.
Let $\chi_1(s)$ and $\chi_2(s)$ be any two linearly independent
solution of the hypergeometric equation
\begin{equation}
\chi''+ \left(\frac{1-\alpha_1}{s}+\frac{1-\alpha_2}{s-1} \right)\chi'+ 
\frac{(\alpha_1+\alpha_2-1)^2-\alpha_3^2}{4s(s-1)}\chi=0 \,.
\label{hyper}
\end{equation}
If the independent variable $t$ in the gDH system is defined by
\begin{equation}
t(s) = \frac{\chi_2(s)}{\chi_1(s)} \,,  
\label{ratio1}
\end{equation}
then the inverse function $s(t)$ satisfies the Schwarzian equation above.
Thus, it is possible to express the gDH variables $\omega_i$
in terms of the hypergeometric solution $\chi_1$ and its derivative.
One should note that $s(t)$ is single-valued if and only if the parameters $\alpha_i$
in (\ref{schwarzian}) and (\ref{hyper}) are either zero or reciprocals of a positive integer.

The reduced system \eqref{Weq} takes a simple but interesting
form if we consider a variable change from $t$ to $s$ and re-express
the corresponding equations. First, let us define new variables
\begin{equation}
\label{What}
\widehat{W_1} = \frac{W_1}{2} \,, \qquad \widehat{W_2} = \frac{W_2}{2i}\,,
\qquad \widehat{W_3} = \frac{W_3}{2i}\,, 
\end{equation}
where $i := \sqrt{-1}$. Then by using the parametrization of the $x_i$ from \eqref{xs} in
\eqref{Weq}, one obtains a non-autonomous, non-homogeneous version
of the Euler ``top'' equations, namely,
\begin{equation}
\label{Whateq}
\begin{split}
\widehat{W_1}'= \frac{\widehat{W_2}\widehat{W_3}}{s-1}+\frac{f(s)}{\sqrt{s-1}} \,,
\qquad
&\widehat{W_2}'= \frac{\widehat{W_1}\widehat{W_3}}{s}-\frac{f(s)}{\sqrt{s}} \,,
\qquad
\widehat{W_3}'= \frac{\widehat{W_1}\widehat{W_2}}{s(s-1)}-\frac{f(s)}{\sqrt{s(s-1)}} \,,
\\
\text{where} \qquad  &f(s) = \frac{\alpha_1^2(s-1)-\alpha_2^2s-\alpha_3^2s(s-1)}{4s(s-1)} \,,
\end{split}
\end{equation}
and ``prime'' indicates derivative with respect to $s$.
It follows from \eqref{Whateq} that
\[\begin{split}
&\widehat{W_1}\widehat{W_1}'-\widehat{W_2}\widehat{W_2}'-\widehat{W_3}\widehat{W_3}' \\
&= \widehat{W_1}\widehat{W_2}\widehat{W_3}
\left(\frac{1}{s-1}-\frac{1}{s}-\frac{1}{s(s-1)}\right)+
f(s)\left(\frac{\widehat{W_1}}{\sqrt{s-1}} + \frac{\widehat{W_2}}{\sqrt{s}}+
\frac{\widehat{W_3}}{\sqrt{s(s-1)}}\right)=0 \,. 
\end{split}  \]
The interested reader can easily verify using \eqref{What}, \eqref{ratio} and
\eqref{W} that the coefficient of $f(s)$ vanishes identically in above,
thereby showing that $\widehat{W_1}^2 - \widehat{W_2}^2-\widehat{W_3}^2$
is a constant.
Moreover, from \eqref{What} one can easily compute 
\[\gamma=\widehat{W_1}^2 - \widehat{W_2}^2-\widehat{W_3}^2
= \frac14 \sum_{i=1}^3W_i^2=\frac14 Q = -\frac14 \,.\]
Thus, the reduced system \eqref{Whateq} for the $\widehat{W_i}$ satisfy
the constraint $\gamma = -\frac 14$.

For the DH case, $f(s)=0$ (because $\alpha_i=0$), then \eqref{Whateq}
reduces to a set of homogeneous, non-autonomous equations arising in similarity reductions of certain
hydrodynamic type systems~\cite{Du1} as well as in self-dual Einstein equations
for $SU(2)$-invariant Bianchi IX metrics~\cite{SChak, Tod, Hi}
(see Remark in Section 2.1). It is known
that this homogeneous system can be solved in terms of a special 
Painlev\'e VI equation via a transformation discussed in \cite{FoAb},
or from the Schlesinger equations associated with the 
Painlev\'e VI equation~\cite{Hi}.
In general, the solution for the reduced system \eqref{Whateq}
can be expressed
in terms of hypergeometric functions utilizing the transformation
given by \eqref{ratio1} and the parametrization of $x_i$ and
$\omega_i$ given in \eqref{xs} and \eqref{omegas}. One also uses
the relation $\dot{s}=1/t'(s) = \chi_1^2/W$ where $\chi_1(s)$ 
is a solution of \eqref{hyper} and $W(s) := W(\chi_1,\chi_2)$
is the Wronskian of two independent solutions.
Finally, taking into account the definitions
from \eqref{W} and \eqref{What} the explicit form of the solutions are
\[\begin{split}
\widehat{W_1}(s) = -\frac{s\sqrt{s-1}}{2}  
&\left(2\frac{\chi_1'}{\chi_1}-\frac{W'}{W}-\frac{1}{s-1}\right)\,, \qquad
\widehat{W_2}(s) = \frac{\sqrt{s}(s-1)}{2}
\left(2\frac{\chi_1'}{\chi_1}-\frac{W'}{W}-\frac{1}{s}\right)\,, \\
&\widehat{W_3}(s) = \frac{\sqrt{s(s-1)}}{2}
\left(2\frac{\chi_1'}{\chi_1}-\frac{W'}{W}-\frac{1}{s}-\frac{1}{s-1}\right)\,,
\end{split}\,.\]
Moreover, applying Abel's formula to \eqref{hyper}, $W'/W$ is expressed as
\[\frac{W'}{W}=-\left(\frac{1-\alpha_1}{s}+\frac{1-\alpha_2}{s-1}\right) \,.\]

A more direct way to solve the $\widehat{W_i}$ is to reduce the system
\eqref{Whateq} into a single, scalar ordinary differential equation
for one of the variables. Recall that the $\widehat{W_i}$ satisfy
the following constraints, namely,
\begin{equation}
\label{constraint}
\widehat{W_1}^2 - \widehat{W_2}^2-\widehat{W_3}^2 = -\frac14 \,,
\qquad
\frac{\widehat{W_1}}{\sqrt{s-1}} + \frac{\widehat{W_2}}{\sqrt{s}}+
\frac{\widehat{W_3}}{\sqrt{s(s-1)}} =0 \,.
\end{equation}
By regarding these constraints as two equations for the 
$\widehat{W_i}$, it is possible to solve for any two of them, say,
$\widehat{W_1}$ and $\widehat{W_3}$ in terms of $\widehat{W_2}$.
Thus, one obtains
\begin{equation}
\label{What1}
\widehat{W_1} =  \frac{c-\sqrt{s}\,\widehat{W_2}}{\sqrt{s-1}}\,, 
\qquad \quad \widehat{W_3} =  \frac{\widehat{W_2} - c\sqrt{s}}{\sqrt{s-1}}\,,
\qquad c=\pm \frac12 \,.
\end{equation}
Next, substituting the expressions for $\widehat{W_1}$ and
$\widehat{W_3}$ from \eqref{What1} into the equation for
$\widehat{W_2}$ in \eqref{Whateq}, yields a Riccatti equation
\[\sqrt{s}(s-1)\widehat{W_2}'+\widehat{W_2}^2-c\,\frac{s+1}{\sqrt{s}}\widehat{W_2}
+(s-1)f(s) + \frac14 =0 \,,\]
where the rational function $f(s)$ is given in \eqref{Whateq}.
If we take $c = \frac 12$, then the Riccatti equation can be linearized 
by the following transformation
\[\frac{\widehat{W_2}}{\sqrt{s}(s-1)} = \frac12\left(\frac{1-\alpha_2}{s-1}
-\frac{\alpha_1}{s}\right) + \frac{\chi'}{\chi} \]
where the function $\chi(s)$ satisfies the hypergeometric equation 
\eqref{hyper}. If $c = -\frac 12$, then one can still linearize the
resulting Riccatti equation but the parameters in the underlying
hypergeometric equation are related to but are not the same 
as the $\alpha_i$.
   
\section{The DH-IX matrix system}
So far we have dealt with the gDH system which consists
of the DH equations together with a common additive term 
$\tau^2$ appearing in all three
equations in \eqref{gendh}. In this section, we will show how
the gDH system can be derived
from a $3 \times 3$ matrix system which arise as a reduction
of the SDYM field equations. We start by reviewing the reduction 
process on the SDYM equations following~\cite{ACH1}.

Consider a gauge group $G$ which may be a finite or infinite-dimensional
Lie group. The gauge field $F$ is a 2-form taking values in the
associated Lie algebra $\mathfrak{g}$, and is given in terms of the
$\mathfrak{g}$-valued connection 1-form (gauge potential) $A$ as
$F=dA-A \wedge A$. In a local co-ordinate system $\{x^a\}\, \, a=0,1,2,3$
the gauge field components are given by 
$F_{ab}=\partial_aA_b-\partial_bA_a - \left[ A_a, A_b \right]$ where $\partial_a$
denotes partial derivative with respect to $x^a$ and $[\,,\,]$ denotes the Lie
bracket in $\mathfrak{g}$. The self-duality condition implies that
$F={^*}F$ where ${^*}F$ is the dual 2-form. In terms of components
of $F$, the self-duality condition is equivalent to   
\begin{equation} \label{sdym} 
F_{0i} = F_{jk}\,, \qquad i\neq j \neq k\,, \,\, cyclic\,. 
\end{equation}
If the connection 1-form is restricted to depend only on
the co-ordinate $x^0 := t$, then without loss of generality, one
can choose a gauge where $A_0 = 0$. Consequently, $A_i = A_i(t)$ 
for $i=1,2,3$, and (\ref{sdym}) reduces to the Nahm equations \cite{Nahm} 
\begin{equation}
\label{nahm}
\dot{A}_i = \left[ A_j, A_k \right]\,, \qquad i\neq j \neq k\,, \,\, cyclic\,.
\end{equation}
Suppose the Lie algebra $\mathfrak{g}$ is chosen to be $\mathfrak{sdiff}(S^3)$ - the
infinite-dimensional Lie algebra of diffeomorphisms on $S^3$ generated
by the left-invariant vector fields $X_i$ satisfying the relation 
$\left[ X_i, X_j \right] = X_k, \,\, i\neq j \neq k\,, \,\, cyclic\,.$ 
Furthermore, let the $A_i$ be of the form  
\begin{equation}
\label{fld} A_i = - \sum_{j,k=1}^3 M_{ij} (t) O_{jk} X_k\, \,\, 
\end{equation}
where $M_{ij}(t)$ are the entries of a $3 \times 3$ matrix $M (t)$ and
$O_{ij} \in SO(3)$ represents a point on $S^3$. Then the 
Nahm equations \eqref{nahm} lead to the following
matrix ordinary differential equation for $M(t)$~\cite{ACH1, Hi1}
\begin{equation}
\label{mateq} \dot{M} = C(M) + M^T M - (\text{Tr}M)M \,,
\end{equation}
where $C(M)$ denotes the matrix of cofactors of $M$. Equation \eqref{mateq}
is a ninth-order coupled system of equations for the matrix elements
of $M(t)$ and was referred to as the DH-IX system in \cite{ACH, ACH1}.
Indeed, by expressing $M$ as
\[
M =
\left({\begin{array}{ccc}
\Omega_1 & \theta_3 & \phi_2 \\
\phi_3 & \Omega_2 & \theta_1 \\
\theta_2 & \phi_1 & \Omega_3
\end{array} } \right)
\]
\eqref{mateq} can be explicitly written out as
\begin{align}
\dot{\Omega}_i &= \Omega_j\Omega_k-\Omega_i(\Omega_j+\Omega_k)
-\theta_i\phi_i + \theta_j^2 + \phi_k^2  \nonumber \\  
\dot{\theta}_i &= -\left(\theta_i+\phi_i\right)\Omega_i-\left(\theta_i-\phi_i\right)\Omega_k 
+ \theta_k \left( \theta_j + \phi_j \right) \label{ind} \\   
\dot{\phi}_i &= -\left(\theta_i+\phi_i\right)\Omega_i+\left(\theta_i-\phi_i\right)\Omega_j 
+ \phi_j \left( \theta_k + \phi_k \right) \,, \nonumber
\end{align}
$i \neq j \neq k = 1,2,3, \, cyclic$.
Equations \eqref{ind} can be regarded as the original DH system 
but with individual additive terms. We next show how to recast the
DH-IX equation into the gDH system \eqref{gendh} where the equations
have a common additive term.

\subsection{Reduction of DH-IX to the gDH system} \label{sec:sec3.1}

Note that the equations for the off-diagonal entries in \eqref{ind} 
involve symmetric and skew-symmetric 
combinations of the off-diagonal elements. This fact can be exploited
further to simplify the matrix equation \eqref{mateq} as follows:
First, the cofactor matrix $C(M) = \left(\mathrm{adj}M\right)^T$, where the 
adjugate matrix can be expressed as
\[\mathrm{adj}M = M^2 - (\mathrm{Tr}M)M + 
\frac12 \left(\left(\mathrm{Tr}M\right)^2 - \mathrm{Tr}M^2\right)I \]
using the Caley-Hamilton theorem for $3 \times 3$ matrices.
In above, $\mathrm{Tr}$ denotes the matrix trace and $I$ is the
identity matrix. Next, substituting the transpose of the above 
expression for $C(M)$ into \eqref{mateq} yields
\begin{equation}
\label{mateq1}
\dot{M} = (M^T-(\mathrm{Tr}M)I)(M+M^T)+
\frac12\left(\left(\mathrm{Tr}M\right)^2-\mathrm{Tr}M^2\right)I \,.
\end{equation}
Equation \eqref{mateq1} motivates decomposing the matrix $M$ into its
symmetric and skew-symmetric parts and re-expressing the 
DH-IX system in terms of these components as illustrated below.
Let us consider the following decomposition of $M$  
\begin{equation}
M = M_s + M_a = P (d + a) P^{-1} \,, 
\label{Mdecomp}
\end{equation}
where the symmetric part $M_s$ is further diagonalized
by a orthogonal matrix $P$ ($P^T=P^{-1}$) and the skew-symmetric
part is expressed as $M_a=PaP^{-1}$ with
\begin{equation} 
 d = \left({\begin{array}{ccc}
\omega_1 & 0 & 0\\
0 & \omega_2 & 0\\
0 & 0 & \omega_3
\end{array} } \right)\,, \qquad  
a = \left({\begin{array}{ccc}
0 & \tau_3 & - \tau_2 \\
- \tau_3 & 0 & \tau_1 \\
\tau_2 & - \tau_1 & 0
\end{array} } \right) \,.
\label{da}
\end{equation}
Substituting \eqref{Mdecomp} into \eqref{mateq1} yields the following set
of matrix equations for $P, a$ and $d$,
\begin{equation}
\dot{P}+Pa = 0\,, \qquad \dot{a}+ad+da=0\,, \qquad 
\dot{d} = 2d^2-2(\mathrm{Tr}d)d+
\frac12\left(\mathrm{Tr}d^2-\left(\mathrm{Tr}d\right)^2-2\mathrm{Tr}a^2\right)I 
\label{mateq2}
\end{equation}
The last equation of \eqref{mateq2} gives the gDH system \eqref{gendh}
with $\tau^2=\tau_1^2+\tau_2^2+\tau_3^2$. Then, using \eqref{xprop1}
one can integrate the second equation in \eqref{mateq2}, i.e.,
\begin{equation}
\label{teq} 
\dot{\tau_i}=-\tau_i(\omega_j+\omega_k) \quad \Rightarrow \quad
\tau_i^2 = \alpha_i^2x_jx_k = 
\alpha_i^2 \left(\omega_j-\omega_i\right)\left(\omega_i-\omega_k\right)\,, 
\quad i \neq j \neq k, \, cyclic\,,   
\end{equation}
and where $\alpha_i$ are integration constants. 
Combining the last equation of \eqref{mateq2} with \eqref{teq},
yields the gDH system \eqref{gendh}. The first equation in
\eqref{mateq2} is linear
and can be solved for $P$ given the $\tau_i$ although it is
not possible to obtain closed form solutions for $P$ except
for special cases. We illustrate one such special case in the
example below.

\subsection{The DH-V system}
We now discuss a fifth order reduction of the DH-IX system where
the matrix $P$ introduced in \eqref{Mdecomp} can be expressed in
closed form. Let us consider the case in which the DH-IX matrix
has the special form
\begin{equation*} 
\label{ca} M = \left({\begin{array}{ccc}
\Omega_1 & \theta & 0 \\
\phi & \Omega_2 & 0 \\
0 & 0 & \Omega_3
\end{array} } \right) \,.
\end{equation*}
Then \eqref{ind} becomes a fifth-order system
given by
\begin{equation} \label{aceq}
\begin{split}
\dot{\Omega}_1 &= \Omega_2 \Omega_3 - \Omega_1 ( \Omega_2 + \Omega_3 ) + \phi^2 \\
\dot{\Omega}_2 &= \Omega_3 \Omega_1 - \Omega_2 ( \Omega_3 + \Omega_1 ) + \theta^2 \\
\dot{\Omega}_3 &= \Omega_1 \Omega_2 - \Omega_3 ( \Omega_1 + \Omega_2 ) - \theta \phi \\
\dot{\theta} &= - \left( \theta + \phi \right) \Omega_3 - \left( \theta - \phi \right) \Omega_2 \\
\dot{\phi} &= - \left( \theta + \phi \right) \Omega_3 + \left( \theta - \phi \right) \Omega_1 
\end{split}
\end{equation}
which was introduced in~\cite{fintgradflo}. We refer to system \eqref{aceq}
as the DH-V system and will construct its solution based on the method
discussed in \hyperref[sec:sec3.1]{Section 3.1}. Due to the special block structure of $M$,
its symmetric part $M_s$ can be diagonalized by an orthogonal matrix
of the form
\begin{equation}
P = \left({\begin{array}{ccc}
\cos \gamma & \sin \gamma & 0 \\
-\sin \gamma & \cos \gamma & 0 \\
0 & 0 & 1
\end{array} } \right)
\label{P}
\end{equation}
where $\gamma=\gamma(t)$ is a complex function to be determined.
That is, $M_s = PdP^{-1}$ with $d$ as in \eqref{da}. Furthermore, the 
skew-symmetric part $M_a$ commutes with the $P$ above so that 
\begin{equation} \label{aeq}
a = P^{-1}M_aP=M_a = \left({\begin{array}{ccc}
0 & \tau_3 & 0 \\
- \tau_3 & 0 & 0 \\
0 & 0 & 0 
\end{array} } \right) \quad \Rightarrow \quad \tau_3 = \frac12(\theta-\phi)
\end{equation}
Since $\tau_1=\tau_2=0$ for the DH-V case, we have $\tau^2=\tau_3^2$
in \eqref{gendh} which is now solved via the Schwarzian equation
\eqref{schwarzian} with $\alpha_1=\alpha_2=0$. Moreover, using
\eqref{teq} and \eqref{xs}, one obtains
\begin{equation}
\label{t3}
\tau_3 = \alpha_3\sqrt{x_1x_2}= \pm \frac{i\alpha_3}{2}\frac{\dot{s}}{\sqrt{s(s-1)}}\,.
\end{equation}
Then the first equation in \eqref{mateq2}, 
\[ \dot{P} = -Pa \qquad \Rightarrow \qquad \dot{\gamma} = \tau_3 \,, \]
which can be solved in terms of $s(t)$ as
\begin{equation}
\label{gamma}
\gamma(s(t)) = \pm i\alpha_3\log\left(\sqrt{s}-\sqrt{s-1}\right) + \gamma_0 \,,
\end{equation}
where $\gamma_0$ is a (complex) constant. Hence the DH-V matrix $M$  
can be reconstructed in terms of the matrices $P, d$ and $a$ as follows:
\begin{equation} \label{transform}
\begin{split}
\Omega_1+\Omega_2=&\omega_1+\omega_2 \,, \qquad 
\Omega_1-\Omega_2 = (\omega_1-\omega_2)\cos 2\gamma \,, \qquad
\Omega_3 = \omega_3\,, \\
&\theta+\phi = (\omega_2-\omega_1)\sin 2\gamma\,,
\qquad \theta-\phi = 2\tau_3\,,
\end{split}
\end{equation}
where $\omega_i$ are given by \eqref{omegas}, and $\tau_3, \gamma$
are given by equations \eqref{t3} and \eqref{gamma}, respectively.
Equation \eqref{transform} gives the complete solution of the 
DH-V system in terms of the solution $s(t)$ of Schwarzian
equation \eqref{schwarzian} with $\alpha_1=\alpha_2=0$.

It is also possible to express the constraint $Q$ introduced 
in Section 2.1 in terms of the DH-V matrix elements. Indeed,
one can calculate from \eqref{transform}
\begin{equation} \label{trule}
\begin{split}
\omega_1 =\frac12\left(\Sigma+\Delta\right)\,, \qquad 
\omega_2 = \frac12\left(\Sigma-\Delta \right)\,, \qquad 
\omega_3 = \Omega_3\,, \\
\Sigma := \Omega_1+\Omega_2\,, \qquad 
\Delta := \pm\sqrt{\left(\Omega_1-\Omega_2\right)^2+\left(\theta+\phi\right)^2}\,.
\end{split}
\end{equation}
Substituting these expressions into the definition of $Q$, yields
\begin{align}
Q  &:= \frac{\omega_1^2}{x_2 x_3} + 
\frac{\omega_2^2}{x_3 x_1} + \frac{\omega_3^2}{x_1 x_2} \nonumber \\
&= \frac{
\frac14\left(\Sigma+\Delta\right)^2\left(\frac12(\Sigma-\Delta)-\Omega_3\right)
-\frac14\left(\Sigma-\Delta\right)^2\left(\frac12(\Sigma+\Delta)-\Omega_3\right)
+\Omega_3^2\Delta}
{\Delta\left(\Omega_3-\frac12(\Sigma+\Delta)\right)\left(\frac12(\Sigma-\Delta)-\Omega_3\right)}
=-1 \nonumber 
\end{align}
after some simplification.

\section{Lax pair and Hamiltonian for the constrained gDH system}
In this section we derive a Lax pair for the reduced non-homogeneous
system \eqref{Whateq} for the $\widehat{W_i}$ introduced in Section 2.2.
Furthermore, we show that \eqref{Whateq} can also be regarded as a 
constrained Hamiltonian system in the phase space of the
variables $\widehat{W_i}$. 

\subsection{Lax equation} 
Specifically, we find $3\times 3$ matrices $U$ and $V$ such that
\eqref{Whateq} is equivalent to the following matrix Lax equation
\[ U' + [U,V]=0 \,,\]
where recall that ``prime'' denotes $\frac{d}{ds}$.
Let us choose $U$ and $V$ in the Lie algebra $so(1,2)$
as follows:
\begin{equation}
\label{lax1} U = \left({\begin{array}{ccc}
0 & \widehat{W_3} &  \widehat{W_2} \\
\widehat{W_3} & 0 &  \widehat{W_1} \\
\widehat{W_2} &  -\widehat{W_1} & 0
\end{array} } \right) \,, \qquad \quad
V = \left({\begin{array}{ccc}
0 & v_3 &  v_2 \\
v_3 & 0 &  v_1 \\
v_2 &  -v_1 & 0
\end{array} } \right) \,,
\end{equation}
where the $v_i$ are to be determined. The commutator $[U,V]$ is also
in $so(2,1)$ and its entries should be equal to the right hand side
of \eqref{Whateq}, which we denote by $r_i, \, i=1,2,3$.
This results in the following linear system
\begin{equation}
\label{Beq}
B{\bm v} = {\bm r}\,, \qquad
B=\left({\begin{array}{ccc}
0 & -\widehat{W_3} &  \widehat{W_2} \\
-\widehat{W_3} & 0 &  \widehat{W_1} \\
\widehat{W_2} &  -\widehat{W_1} & 0
\end{array} } \right)\,, \qquad {\bm v} = [v_1, v_2, v_3]^T\,,
\quad {\bm r}=[r_1, r_2, r_3]^T \,.
\end{equation} 
for the vector ${\bm v}$. Note that the matrix $B$ is singular. 
In order for the linear system \eqref{Beq} to have a {\it consistent}
solution, the vector ${\bm r}$ must be orthogonal to the
null space of $B^T$ by Fredholm's alternative.
The null space of $B^T$ is spanned by the vector 
$\widehat{\bm N} = [\widehat{W_1}, -\widehat{W_2}, -\widehat{W_3}]^T$.
Therefore, one must have $\widehat{\bm N} \bm{\cdot r} =
\widehat{W_1}r_1-\widehat{W_2}r_2-\widehat{W_3}r_3=0$, which 
is readily verified from the calculations immediately following \eqref{Whateq}.
Thus, the linear system \eqref{Beq} admits infinitely many
solutions (defined modulo the homogeneous solution spanned by
the null vector $[\widehat{W_1}, \widehat{W_2}, \widehat{W_3}]^T$
of $B$). A particular choice for the vector ${\bm v}$ is given by
\[ v_1=0\,, \qquad v_2=-\frac{r_3}{\widehat{W_1}} =
-\left(\frac{\widehat{W_2}}{s(s-1)} + \frac{f(s)}{\widehat{W_1}\sqrt{s(s-1)}}\right)\,,
\qquad  v_3=\frac{r_2}{\widehat{W_1}} =
\left(\frac{\widehat{W_3}}{s} + \frac{f(s)}{\widehat{W_1}\sqrt{s}}\right)\,,\]
which then yields the matrix $V$ in the Lax pair.

In a general setting, the Lax equation $U'+[U,V]=0$ is useful
to generate a sequence of conserved quantities $\mathrm{Tr} U^n, \, n=1,2,\ldots$.
Indeed, by differentiating with respect to $s$ one obtains
\[\left(\mathrm{Tr} U^n\right)' = n\mathrm{Tr}\left(U^{n-1}[V,U]\right) 
=n\mathrm{Tr}\left(V[U,U^{n-1}]\right)=0 \,.\]
These conserved quantities are related to the symmetric functions of 
the eigenvalues of of the matrix $U$. In the present
case, the eigenvalues of $U$ are simply given by 
$\lambda = 0, \pm\sqrt{- \bm q} = 0, \pm \frac12$ since  
${\bm q} =\frac14 Q = -\frac14$. In fact, one obtains $\mathrm{Tr}U=0,
\mathrm{Tr}U^2 = -2 {\bm q}$, and the remaining traces are polynomials
in $\bm q$.      

It is worth pointing out that \eqref{Weq} for the $W_i$
also admits a Lax pair. Here, one chooses $so(3)$-valued 
$3 \times 3$ matrices
\begin{equation}
\label{lax2} L = \left({\begin{array}{ccc}
0 & W_3 &  -W_2 \\
-W_3 & 0 &  W_1 \\
W_2 &  -W_1 & 0
\end{array} } \right) \,, \qquad \quad
A = \left({\begin{array}{ccc}
0 & A_3 &  -A_2 \\
-A_3 & 0 &  A_1 \\
A_2 &  -A_1 & 0
\end{array} } \right) \,,
\end{equation}
where the $W_i(t)$ are defined in \eqref{W} and the $A(t)$
are to be determined such that the Lax equation $\dot{L}+[L,A]=0$  
is equivalent to the system \eqref{Weq}. The matrix $A$ can be found
by proceeding in a similar fashion as outlined above. One finds that
a particular choice for the matrix elements of $A$ is given by
\[A_1=0\,, \qquad A_2=x_3W_2 + \frac{\tau^2}{W_1\sqrt{x_1x_2}} \,,
\qquad A_3=-\left(x_2W_3 + \frac{\tau^2}{W_1\sqrt{x_1x_3}} \right) 
\,.\]
The eigenvalues of $L$ is given by $\lambda=0, \pm\sqrt{-Q} = 0, \pm1$.
Consequently, $\mathrm{Tr}L^n, \, n=1,2,\ldots$ are polynomials 
in $Q$.

\subsection{Hamiltonian formulation}
Equations \eqref{Whateq} can be also regarded as a constrained
Hamiltonian system in the phase space of the variables
$\widehat{W_i}$ satisfying the constraints in \eqref{constraint}.
The phase space is endowed with a natural Poisson structure 
inherited from the Lie-Poisson structure defined on the dual space
of the Lie algebra $so(1,2)$ used to construct the Lax pair.
Explicitly, the Poisson structure is given by the fundamental
Poisson bracket relations
\begin{equation}
\label{pb}
\{\widehat{W_1}, \widehat{W_2}\} = \widehat{W_3} \,, \qquad 
\{\widehat{W_2}, \widehat{W_3}\}=-\widehat{W_1}\,, \qquad 
\{\widehat{W_3}, \widehat{W_1}\}= \widehat{W_2} \,. 
\end{equation}
In general, the Poisson bracket of any two continuously 
differentiable functions $f$ and $g$ on the phase space, is given by
\[\{f,g\} = J(df, dg) = \sum_{i,j,k=1}^3 C_{ij}^k\widehat{W_k}
\frac{\partial f}{\partial \widehat{W_i}}\frac{\partial f}{\partial\widehat{W_j}}
\,, \]
where $C_{ij}^k$ are the structure constants for the Lie algebra $so(1,2)$.
The Poisson tensor $J_{ij} := \displaystyle \sum_{k=1}^3C_{ij}^k\widehat{W_k}$
is degenerate on the three-dimensional phase space, and admits a Casimir 
function constructed from the Lax matrix $U$ as follows  
\[ C = -\frac12 \mathrm{Tr}U^2 = \widehat{W_1}^2-\widehat{W_2}^2-\widehat{W_3}^2\]
such that $J(\cdot, dC) = 0$. In other words, $\{f,C\}=0$ for any smooth
function $f$ on the phase space. Note from \eqref{constraint} that $C+\frac 14 =0$
is one of the constraints, while the other constraint is given by $l=0$, where
\[l = -\frac12 \mathrm{Tr}US = \frac{\widehat{W_1}}{\sqrt{s-1}} + 
\frac{\widehat{W_2}}{\sqrt{s}}+
\frac{\widehat{W_3}}{\sqrt{s(s-1)}} \, \]

\[S = \left({\begin{array}{ccc}
0 & (s(s-1))^{-1/2} &  s^{-1/2} \\
(s(s-1))^{-1/2} & 0 &  (s-1)^{-1/2} \\
s^{-1/2}  &  -(s-1)^{-1/2} & 0
\end{array} } \right) \,. \]
Next, we introduce a Hamiltonian function on the phase space by
\begin{equation}
\label{H}
H = -\frac 12 \mathrm{Tr}(UIU -4cf(s)US) =
\frac 12 \left(\frac{\widehat{W_1}^2}{s-1}-\frac{\widehat{W_2}^2}{s}
- \frac{(2s-1)\widehat{W_3^2}}{s(s-1)}\right) -4cf(s)l  \,, \\
\end{equation}
where $I=\mathrm{diag}(s^{-1}, (s-1)^{-1}, 0)$, $l$ is defined above, 
$c=\pm \frac12$, and $f(s)$ is defined in \eqref{Whateq}. 
With the fundamental Poisson brackets given by \eqref{pb}, the reduced
gDH equations \eqref{Whateq} can be expressed by the following
equation of motions together with the constraints
\begin{equation}
\label{Heq} 
\dot{\widehat{W_i}} = \{\widehat{W_i}, H \}\,, \quad \qquad
C+\frac 14=0\,, \quad \qquad l = 0 \,,
\end{equation}
where the Hamiltonian $H$ is given by \eqref{H}. The equations of 
of motions obtained from \eqref{Heq} determines the equations
in \eqref{Whateq} after applying the constraints. For example,
one can compute using \eqref{pb} that
\[\{\widehat{W_1}, H \} = \frac{\widehat{W_2}\widehat{W_3}}{s-1}   
-4cf(s)\left(\frac{\widehat{W_2}}{\sqrt{s(s-1)}} - 
\frac{\widehat{W_3}}{\sqrt{s}}\right) \,.\]
Upon applying the constraints, one can replace $\widehat{W_3}$ in the 
second term above by its expression from \eqref{What1} to
obtain the first equation in \eqref{Whateq}. The remaining equations
in \eqref{Heq} lead to the corresponding equations
in \eqref{Whateq} in a similar fashion.

For consistency, it also needs to be checked that the constraints
are satisfied by the Hamiltonian dynamics. In other words, 
one should have modulo the constraints
\[ \frac{dC}{ds} = \{C, H\}=0\,, \qquad \qquad \frac{dl}{ds} = 
\frac{\partial l}{\partial s} + \{l, H\} = 0 \,.\]
The first consistency condition is obviously satisfied
since $C$ is a Casimir function, the second one can also be verified
by using \eqref{What1} and after some straightforward computations.

\section{Conclusion}
In this note, we have discussed certain features pertaining to the
integrability of the gDH system introduced in \cite{ACH, ACH1}
which contains the well-known DH system as a special case.
We have also provided a detailed derivation of the gDH system from
a ninth-order system which arises as a reduction of the SDYM
equations associated with Lie algebra $\mathfrak{sdiff}(S^3)$.
Starting from the gDH system, we have derived a reduced system 
\eqref{Whateq} which is similar to a non-autonomous, non-homogeneous Euler
system with a constraint. We then give the complete solution
for this system in terms of the general solution of the classical 
hypergeometric equations. Finally, we derive a Lax pair and
a Hamiltonian for this reduced system. 

\section*{Acknowledgement}
The work of SC was partially supported by NSF grant No. DMS-1410862 and the
work of PG was partially supported by FAPESP through IFSC, Sao Carlos with 
grant number 2016/06560-6. SC also thanks SNBNCBS, Kolkata, for its hospitality 
where this work was initiated.
PG would like to express his gratitude to Professor Maxim Kontsevich and other
members of the IHES for their warm hospitality where the final part of the work has been done.

\end{document}